\documentclass[twocolumn,runningheads,natbib]{svjour2}
\smartqed  
\usepackage{graphicx}
\usepackage{amsmath}
\usepackage{amssymb}
\newcommand{\psr}{PSR J1119--6127}
\newcommand{\chan}{Chandra}
\newcommand{\xmm}{XMM--Newton}
\newcommand{\snr}{G292.2--0.5}
\journalname{Astrophysics and Space Science}
\begin{document}

\title{PSR J1119--6127 and the X-ray Emission from High Magnetic Field Radio Pulsars
}


\author{M. E. Gonzalez, V. M. Kaspi, F. Camilo, B. M. Gaensler and M. J. 
Pivovaroff}

\authorrunning{Gonzalez et al.} 

\institute{M. E. Gonzalez and V. M. Kaspi \at
              Department of Physics, Rutherford Physics Building, 
McGill University, Montreal, QC H3A 2T8, Canada. 
              \email{gonzalez@physics.mcgill.ca}           
           \and
           F. Camilo \at
            Columbia Astrophysics Laboratory, Columbia University, 550 West 
120th Street, New York, NY 10027, USA.
           \and
           B. M. Gaensler \at
          Harvard-Smithsonian Centre for Astrophysics, 60 Garden Street, 
Cambridge, MA 02138, USA; Current address: School of Physics, The University of Sydney, NSW 2006, 
Australia.         
	\and
           M. J. Pivovaroff \at
            Lawrence Livermore National Laboratory, P.O. Box 808, L-258, 
Livermore, CA 94550, USA.
}

\date{Received: date / Accepted: date}

\maketitle

\begin{abstract}
The existence of radio pulsars having inferred magnetic fields in the magnetar regime 
suggests that possible transition objects could be found in the radio pulsar population. The 
discovery of such an object would contribute greatly to our understanding of neutron star physics. 
Here we report on unusual X-ray emission detected from the radio pulsar PSR J1119--6127 
using XMM--Newton. The pulsar has a characteristic age of 1,700 yrs and inferred surface 
dipole magnetic field strength of 4.1$\times$10$^{13}$ G. In the 0.5-2.0 keV range, the emission
shows a single, narrow pulse with an unusually high pulsed fraction of $\sim$70\%. No pulsations 
are detected in the 2.0--10.0 keV range, where we derive an upper limit at the 99\% level for the 
pulsed fraction of 28\%. The pulsed emission is well described by a thermal blackbody model with 
a high temperature of $\sim$2.4$\times$10$^{6}$ K. While no unambiguous signature of 
magnetar-like emission has been found in high-magnetic-field radio pulsars, the X-ray 
characteristics of PSR J1119--6127 require alternate models from those of conventional 
thermal emission from neutron stars. In addition, PSR J1119--6127 is now the radio pulsar 
with the smallest characteristic age from which thermal X-ray emission has been detected. 

\end{abstract}

\section{Introduction}\label{intro}
The emission from the $\sim$1,500 radio pulsars  (PSRs) discovered to date is generally 
thought to be powered by the loss of rotational kinetic energy due to magnetic braking. 
Radio pulsars with implied magnetic fields in the range 
$\sim$10$^{13-14}$~G have now been discovered, showing that radio emission can be
produced in neutron stars with fields above the quantum critical field $B_c$ = 
4.4$\times$10$^{13}$~G. More exotic neutron stars with magnetic fields in the range 
$\sim$10$^{14-15}$~G have been observed at high energies and are believed to be powered 
by the decay of their large magnetic fields. Anomalous X-ray pulsars (AXPs) and Soft-Gamma 
Repeaters (SGRs) make up this class of so-called ``magnetars" \citep{wt04}. As the inferred 
magnetic field strengths 
of radio pulsars and magnetars are now found to overlap, the underlying physical reasons for the
differences in their emission properties remains a puzzle. To date, no radio pulsar has been
observed to exhibit magnetar-like emission at high energies, posing an interesting challenge to 
current emission theories.

Here we summarize the analysis and results obtained from an \xmm\ observations of \psr, 
originally published by \citet{gkc+05}. \psr\ is one of the youngest radio pulsars known and also
has one the highest inferred magnetic fields in the radio pulsar population. Our observation 
reveals unusual thermal emission from this object which may represent the first evidence for 
high-magnetic-field effects in the emission from a ``normal'' radio pulsar. We discuss these 
results in light of recent theoretical work on emission from highly magnetized neutron stars 
and observations of similar sources.

\psr, with spin period $P=0.408$~s, is among the youngest radio 
pulsars known \citep{ckl+00}. The measured braking index for the pulsar
of $n$ = 2.91$\pm$0.05 ($\dot{\nu}$ $\propto$ --$\nu^{n}$) implies an upper
limit for the age of 1,700~yr. The pulsar
has spin-down luminosity $\dot{E}$ $\equiv$ 4$\pi^2$$I$$\dot{P}$/$P^{3}$ = 
2.3$\times$10$^{36}$~erg~s$^{-1}$ (for a moment of inertia $I$ = 
10$^{45}$~g~cm$^{2}$) and an inferred surface dipole magnetic field 
strength at the equator of $B$ $\equiv$ 3.2$\times$10$^{19}$($P$$\dot{P}$)$^{1/2}$~G = 
4.1$\times$10$^{13}$~G. 
This value of $B$ is among the highest known in the radio pulsar 
population. \psr\ powers a small ($3''$$\times$$6''$) X-ray 
pulsar wind nebula \citep[PWN, ][]{gs03}, which results from the 
confinement of the pulsar's relativistic wind of particles and electromagnetic fields by the 
ambient medium . The pulsar lies close to the center of a $15'$-diameter ``shell'' supernova 
remnant (SNR), \snr\ \citep{cgk+01b,pkc+01}, located at a 
distance of 8.4$\pm$0.4~kpc, as determined from neutral 
hydrogen absorption measurements \citep{cmc04}.

\section{Observation}\label{sec:obs}
We observed \psr\ using the European Photon Imaging Camera (EPIC) 
aboard the \xmm\ satellite on June 26, 2003. The MOS 
and PN instruments were operated in full-window and large-window mode,
respectively. The temporal 
resolution was 2.6~s for MOS and 48~ms for PN. The data 
were analyzed using the Science Analysis System software 
(SAS v6.2.0) and standard reduction techniques. 
The effective exposure time was 48~ks for MOS1/MOS2 and 43~ks for PN. 

\begin{figure}
\centering
  \includegraphics[width=8.2cm]{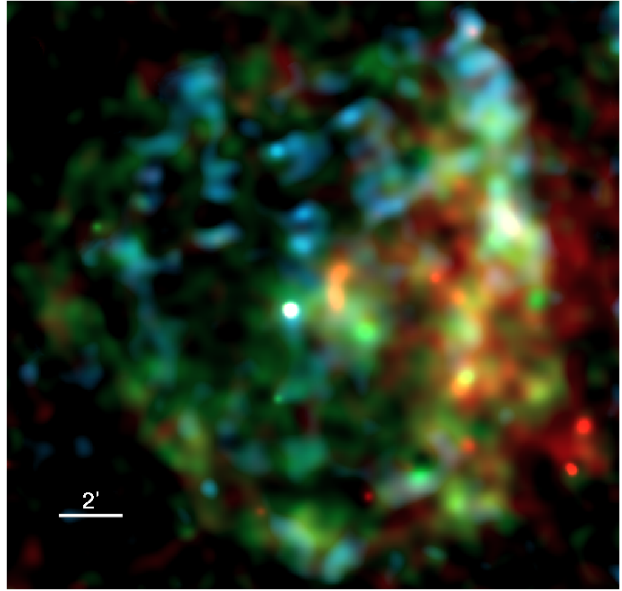}
\caption{Combined MOS image of \snr\ and \psr\ in the 0.3--1.5 keV 
({\it red}), 1.5--3.0 keV ({\it green}) 
and 3.0--10.0 keV ({\it blue}) bands.
Individual MOS1/MOS2 images were first binned into pixels of 
2.5$''$$\times$2.5$''$ and unrelated point sources in the field were 
excluded. These images were then added and adaptively 
smoothed with a Gaussian with $\sigma$=5$''$--15$''$ to obtain 
signal-to-noise ratios higher than 3. Background images and exposure 
maps at each energy band were similarly obtained and used to correct 
the final images.}
\label{FigMOS}       
\end{figure}

Figure \ref{FigMOS} shows the combined MOS image of the 
system in the 0.3--1.5 keV ({\it red}), 1.5--3.0 keV ({\it green}) 
and 3.0--10.0 keV ({\it blue}) bands\footnote{The detailed spatial distribution in the 
3.0--10.0 keV ({\it blue}) image should be examined with caution as this 
energy band suffered from a high degree of stray-light contamination
from a nearby high-energy source.}. The bright source at the center 
has coordinates $\alpha_{2000}$=11$^{h}$19$^{m}$14.65$^{s}$ 
and $\delta_{2000}$=--61$^{\circ}$27$'$50.2$''$ (4$''$ error). This
position coincides with the \chan\ and radio coordinates of \psr. 
Although the 
spatial resolution of \xmm\ (half power diameter of 15$''$) 
permits the PWN to be neither resolved nor separated from 
the pulsar emission itself, \xmm's sensitivity and high time 
resolution allow us to separate the pulsar's emission spectrally and 
temporally. The image reveals for the first
time the detailed morphology of the SNR at X-ray energies as \xmm's 
sensitivity was needed due to the remnant's 
low surface brightness. The large east-west asymmetry at low energies has been 
attributed to the presence of a molecular cloud on the east side of the 
field \citep{pkc+01}.

\section{Results}\label{sec:res}

\subsection{Timing Analysis}\label{ssec:tim}
The PN data were used to search for pulsations from \psr. 
A circular region of 25$''$-radius centered at the X-ray 
coordinates of the pulsar was used to extract the source photons.
The data were divided into 
different energy ranges at 0.5--10.0 keV (620$\pm$30 photons), 
0.5--2.0 keV (340$\pm$25 photons) and 2.0--10.0 keV (275$\pm$22 photons).
The most significant signal was detected in the 0.5--2.0 keV range with 
$\textit{Z}^{2}_{2}$~= 52.8 (6.6$\sigma$ significance) at a frequency of 
2.449726(6) Hz (1$\sigma$ errors). This frequency is in agreement with 
the radio prediction for \psr\ obtained using regular monitoring with
the Parkes telescope. 
In the 2.0--10.0 keV range, no signal was found with a significance 
$>$2.8$\sigma$. In the 0.5--10.0 keV 
range, the above signal was detected with a 5.2$\sigma$ significance.

Figure \ref{FigProfPha} shows the resulting pulse 
profiles at 0.5--2.0 keV (top, left) and 2.0--10.0 keV (top, right). 
The background was estimated from a nearby region away from 
bright SNR knots. The horizontal dashed lines represent our estimates for 
the contribution from the pulsar's surroundings 
obtained using the \chan\ observation of the pulsar \citep{gs03}. The \chan\
count rate excluding the pulsar was used to estimate the \xmm\ counts using 
{\it WebPIMMS}. The resulting pulsed fraction [PF $\equiv$ 
($F_{max}$--$F_{min}$) / ($F_{max}$+$F_{min}$)] is labeled in Figure 
\ref{FigProfPha} (1$\sigma$ statistical errors). In the 2.0--10.0 keV 
range, we derive an upper limit for the pulsed fraction of 28\% \citep[at the 99\%
confidence level; see, e.g.,][]{vvw+94,rem02}. The radio emission from \psr\ 
consists of a single peak of duty cycle 5\% and luminosity at 1.4~GHz of 
28~mJy~kpc$^2$ \citep{ckl+00}. Phase zero in the X-ray profiles corresponds to 
the radio peak (determined with 3~ms uncertainty). Taking into account 
the low temporal resolution of the \xmm\ observation (phase bin width of 51 
ms), this result suggests that the radio peak is in phase with the X-ray 
peak within our uncertainties, or possibly just slightly ahead. Additional
observations at higher temporal resolution will help to confirm and constrain these
results.

\subsection{Spectral Analysis}\label{ssec:spec}
The EPIC data were used to perform a spectral analysis of \psr. 
Circular regions with radii of 20$''$ and 25$''$ were used for MOS and 
PN, encompassing $\sim$75\% and $\sim$78\% of the source photons, 
respectively. The derived fluxes have been corrected accordingly.  
Background regions were chosen from nearby areas away from bright SNR 
knots. The spectra were fit in the 0.5--10.0 keV range using
XSPEC (v.11.3.0) with a minimum of 20 counts per bin from a total of 
240$\pm$19, 210$\pm$18, and 620$\pm$30 background-subtracted counts in 
MOS1, MOS2 and PN, respectively.

Two-component models were needed in order 
to describe the low and high energy portions of the spectra.
A non-thermal power-law component with photon 
index $\Gamma$ $\sim$ 1.5 described the high-energy emission in the spectra 
well. In turn, various models were used to describe the low-energy emission.
The derived fits are summarized in Table \ref{tabAvg}.

%
\begin{table}[h]
\caption{Fits to the \xmm\ phase-averaged spectrum of \psr}
\centering
\label{tabAvg}       
\begin{tabular}{lccc}
\hline\noalign{\smallskip}
& PL+PL & BB+PL & Atm$^{a}$+PL  \\
Parameter  &($\pm$1$\sigma$) & ($\pm$1$\sigma$) & ($\pm$1$\sigma$)\\[3pt]
\tableheadseprule\noalign{\smallskip}
$N_{H}$ (10$^{22}$~cm$^{-2}$) & 2.3$^{+0.4}_{-0.3}$ & 1.6$^{+0.4}_{-0.3}$ & 1.9$^{+0.5}_{-0.3}$ \\
$\chi^2$(dof) & 79(66) & 78(66) & 78(66) \\ [3pt]
& \multicolumn{3}{c}{\it Soft component characteristics} \\ 
$\Gamma$ or $T^{\infty}$ & 6.5$\pm$0.9 & 2.4$^{+0.3}_{-0.2}$~MK &  0.9$\pm$0.2~MK \\
$R^{\infty}$ (km) & ... & 3.4$^{+1.8}_{-0.3}$ & 12 (fixed) \\
$d$ (kpc) & 8.4 (fixed) & 8.4 (fixed) & 1.6$^{+0.2}_{-0.9}$ \\
$f_{abs}$$^{b}$ (10$^{-14}$) & 2.1$^{+2.3}_{-0.9}$ & 1.5$^{+1.8}_{-0.2}$ & 1.7$^{+7.0}_{-0.4}$ \\
$f_{unabs}$$^{b}$ (10$^{-13}$) & 63$^{+57}_{-32}$ & 2.4$^{+3.0}_{-0.5}$ & 7.2$^{+31}_{-1.6}$ \\
$L_{X}$$^{b}$ (10$^{33}$) & 53$^{+50}_{-27}$ & 2.0$^{+2.5}_{-0.4}$ & 0.22$^{+0.88}_{-0.05}$\\  [3pt]
& \multicolumn{3}{c}{\it Hard component characteristics} \\ 
$\Gamma$ & 1.3$^{+0.5}_{-0.2}$ & 1.5$^{+0.3}_{-0.2}$   & 1.5$^{+0.2}_{-0.3}$ \\
$f_{abs}$$^{b}$ (10$^{-14}$) & 7.1$^{+10}_{-1.5}$   & 7.4$^{+3.6}_{-1.0}$ & 7.3$^{+4.7}_{-2.7}$ \\
$f_{unabs}$$^{b}$ (10$^{-13}$) & 1.0$^{+1.6}_{-0.2}$  & 1.1$^{+0.6}_{-0.2}$ & 1.1$^{+0.8}_{-0.3}$ \\
$L_{X}$$^{b}$ (10$^{33}$) & 0.8$^{+1.3}_{-0.2}$ & 0.9$^{+0.5}_{-0.1}$ & 0.04$\pm$0.02 \\ 
\noalign{\smallskip}\hline
\end{tabular}
\flushleft $^{a}$ The atmospheric model was computed with $B$=10$^{13}$~G and pure 
hydrogen composition. The local values for the temperature, $T$, and radius, 
$R$ = 10~km, of the star have been redshifted to infinity according to the formulae
$T^{\infty}$=$T$(1--2$G$$M$/$R$$c^{2}$)$^{1/2}$ 
and $R^{\infty}$=$R$(1--2$G$$M$/$R$$c^{2}$)$^{-1/2}$, with $M$ = 1.4~$M_{\odot}$. \\[-1pt]
\flushleft $^{b}$ The 0.5--10.0 keV absorbed and unabsorbed fluxes, $f_{abs}$ 
and $f_{unabs}$, have units of ergs~s$^{-1}$~cm$^{-2}$. The 0.5--10.0 keV X-ray 
luminosity, $L_X$, at the distance $d$, is in units of ergs~s$^{-1}$. \\
\end{table}

\begin{figure*}[ht]
\centering
  \includegraphics[width=0.75\textwidth]{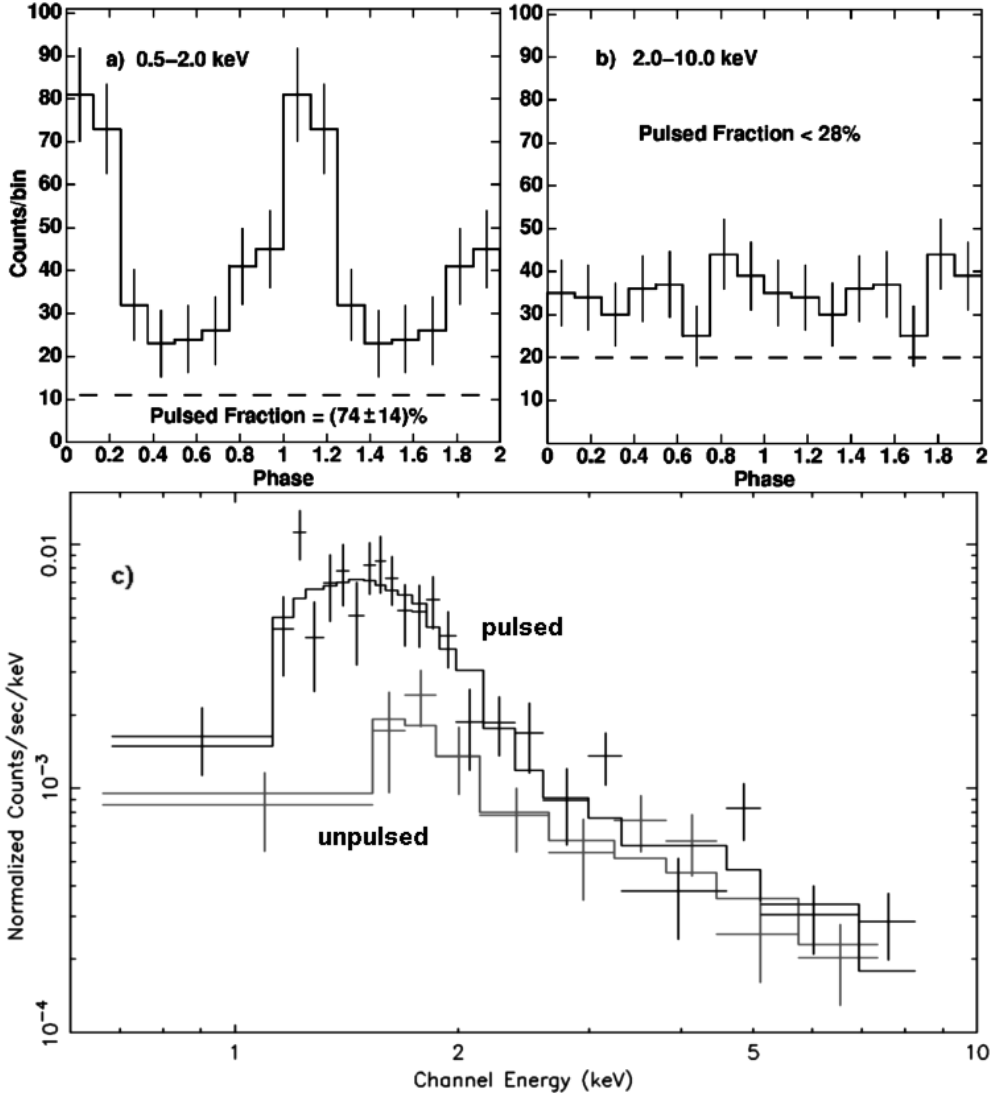}
\caption{{\it Top}: X-ray pulse profiles of PSR \psr\
in the 0.5--2.0~keV ({\it left}) and 2.0--10.0 keV ({\it right}) ranges.
Errors bars are 1$\sigma$ and two cycles are shown. The dashed lines represent 
our estimates for the 
contribution from the pulsar's surroundings (see \S\ref{ssec:tim}). 
{\it Bottom}: EPIC-PN spectra obtained for the pulsed 
($black$) and unpulsed ($red$) regions of the pulse profile with their 
respective best-fit blackbody plus power-law model (solid curves).}
\label{FigProfPha}       
\end{figure*}

We also extracted PN spectra from the ``pulsed'' and ``unpulsed'' regions 
of the pulse profile, at phases 0.7--1.3 (430$\pm$28 counts) and 0.3--0.7 
(200$\pm$20 counts), respectively. These spectra are shown in Figure 
\ref{FigProfPha} (bottom) and were well fit by two-component models that 
agree with those 
derived for the phase-averaged spectra. For example, a blackbody plus power-law
model yielded $T^{\infty}$ = 2.8$\pm$0.4~MK and $\Gamma$ = 1.4$^{+0.5}_{-0.2}$ 
(1$\sigma$ errors). The main difference between the pulsed and unpulsed spectra 
was found to be the relative contributions of the model components.  The pulsed 
spectrum is dominated by the soft component below $\sim$2~keV, while the 
unpulsed spectrum is dominated by the hard, power-law component at 
all energies.

%

\section{Discussion}\label{sec:disc}

\subsection{Observed Emission Characteristics}\label{ssec:emis}
The pulse profile in the 
0.5--2.0 keV range shows a single, narrow pulse with a high pulsed fraction.
Modeled as a Gaussian, the full-width at
half maximum is 0.26$^{+0.08}_{-0.06}$$P$ (1$\sigma$ errors) with 
$\chi^{2}$(dof) = 2.1(4) and a probability of 0.72. A sinusoidal 
fit resulted in $\chi^{2}$(dof) = 8.9(5) with a probability of 
0.11. Due to the limited statistics available, the Gaussian fit is preferred only 
at the 2.4$\sigma$ level (according to the $F$-test). A two-component profile 
cannot be ruled out with the present data (e.g., sine curve plus a narrow peak, or 
two narrow peaks at phases $\sim$0.8 and $\sim$1.1). Additional X-ray observations 
at higher temporal resolution are needed to further constrain the pulse shape.

The observed spectrum from \psr\ requires two-component models. The hard,
non-thermal component is consistent with arising from the pulsar's surroundings.
Using the high-resolution \chan\ data we estimate that the PWN plus SNR emission 
within a $25''$ radius, excluding the pulsar, is well described by a power-law model 
with $\Gamma$=1.8$^{+0.8}_{-0.6}$ and unabsorbed flux in the 0.5--10.0 keV 
range of 0.9$^{+0.6}_{-0.5}$$\times$10$^{-13}$ erg~s$^{-1}$~cm$^{-2}$ 
(1$\sigma$ errors). These values are in good agreement with the hard 
power-law component shown in Table \ref{tabAvg}. Although a small contribution 
from the pulsar to this hard emission cannot be ruled out, it would not 
affect our results on the pulsar's soft emission.

The soft spectral component must then arise from the pulsar.
Non-thermal X-ray spectra from radio pulsars have photon indices in the 
range 0.5$<$$\Gamma$$<$2.7 \citep{ba02,pzs02}. Models for synchrotron 
emission in the pulsar magnetosphere\citep{cz99,rd99} predict $\Gamma$ 
$\lesssim$ 2. The pulsed emission we detect from \psr, if interpreted 
as non-thermal in origin, has a steeper spectrum ($\Gamma$=6.5$\pm$0.9, 
Table~1) than those observed or predicted. Therefore, 
a thermal origin for the observed emission is strongly favored with the
present data. In addition, atmospheric models require small distances (or 
conversely, implausibly large emitting radii at 8.4~kpc) to account for the 
observed emission. We therefore favor a blackbody model to account 
for the observed emission.

The blackbody temperature of the X-ray emission from \psr\ is 
$T_{bb}^{\infty}$ = (2.4$^{+0.3}_{-0.2}$)$\times$10$^{6}$~K. 
This temperature is among the highest seen in radio 
pulsars; while it is naively similar to those found in much older pulsars, 
none of them exhibits a higher temperature at statistically significant 
levels \citep[e.g., PSR J0218+4232 has a characteristic age of $\tau_{c}$ = 
$P$/2$\dot{P}$ = 5.1$\times$10$^{8}$~yr and a blackbody 
temperature of $T_{bb}^{\infty}$ = 2.9$\pm$1.1 
$\times$10$^{6}$~K, 3$\sigma$ errors, ][]{wob04}. \psr\ is now the youngest 
radio pulsar from which thermal emission has been detected, the next 
youngest being Vela ($\tau_{c}$ = 11~kyr) with a blackbody temperature 
$T_{bb}^{\infty}$ = (1.47$\pm$0.18)$\times$10$^{6}$~K 
\citep[3$\sigma$ range, ][]{pzs+01b}. Moreover, the pulsed fraction of the thermal 
emission from \psr\ is significantly higher than is seen in 
other radio pulsars; thermal sources where the emission arises from 
the entire surface or for localized regions show pulsed fractions 
at low energies of $\lesssim$40\% \citep{pzs+01b,pzs02,ba02}. The narrow peak 
in the pulse profile points to yet another difference from what is 
normally seen in thermal emission from radio pulsars at low 
energies \citep{pzs02,ba02}, namely broad pulsations. 

%
\begin{table*}[t]
\centering
\caption{High-Magnetic Field Radio Pulsars}     
\label{tabPsrs}  
\begin{tabular}{ l p{2.0cm}  p{2.2cm}  p{2.2cm}  p{2.2cm}  p{2.0cm}  p{2.0cm} }
\hline\noalign{\smallskip}
PSR & J1847--0130 & J1718--3718 & J1814--1744 & J1846--0258 & B0154--61 & B1509--58\\
\tableheadseprule\noalign{\smallskip}
P (sec) & 6.7 & 3.4 & 4 & 0.32 & 2.35 & 0.15\\
$B$ (10$^{13}$ G) & 9.4 & 7.4 &5.5 & 4.8 & 2.1 & 1.5\\ 
$\tau_c$ (kyr) & 83 & 34 & 85 & 0.72 & 197 & 1.7\\
$\dot{E}$ (ergs s$^{-1}$) & 1.7$\times$10$^{32}$ & 1.5$\times$10$^{33}$ & 4.7$\times$10$^{32}$ & 8$\times$10$^{36}$ & 5.7$\times$10$^{32}$ & 1.8$\times$10$^{37}$ \\
$D$ (kpc) & $\sim$8 & 4--5 &$\sim$10 & $\sim$19 & $\sim$1.7 & $\sim$5 \\
$L_X$ (ergs s$^{-1}$) & $<$5$\times$10$^{33}$ & $\sim$10$^{30}$ & $<$6$\times$10$^{35}$ & 6.4$\times$10$^{34}$ & $<$1.4$\times$10$^{32}$ & 2.4$\times$10$^{34}$\\
$T$ or $\Gamma$  & -- & $T$$\sim$1.6 MK & -- & $\Gamma$$\sim$1.4  & -- & $\Gamma$$\sim$1.4  \\
Ref. & \cite{msk+03} & \cite{km05} &  \cite{pkc00} &  \cite{hcg03} &  \cite{gklp04} &  \cite{gak+02}\\ 
\noalign{\smallskip}\hline
\end{tabular}
\end{table*}

\subsection{Thermal Emission Mechanisms} \label{ssec:mech}
Conventional models for thermal emission from neutron stars cannot account for the observed
characteristics in \psr. Thermal emission from polar-cap 
reheating has been well studied and, whether the required return currents 
arrive from the outer gap region \citep{cz99} or from close to the polar 
cap \citep{hm01b}, the X-ray luminosity is constrained to be $\lesssim$10$^{-5}$$\dot{E}$ 
for sources as young as \psr. This is at least 2 orders of magnitude below what we 
observe.

Thermal emission may also arise from the surface due to initial cooling. The observed 
luminosity is consistent with predictions from standard models of cooling 
neutron stars \citep{ygk+04,pgw05}. However, the effective blackbody temperature is higher 
than predicted and the observed blackbody radius is smaller than allowed from neutron 
star equations of state \citep{lp00b}. The very high observed pulsed fraction is also consistent
with emission arising from a small fraction of the neutron star surface.

On the other hand, recent work on surface emission from highly magnetized neutron stars 
has explored the effects of a high magnetic field, as 
heat conductivity is expected to be suppressed perpendicular to the field lines and will be 
channeled along the lines instead \citep{gkp04,pmp06}. This will produce a highly 
anisotropic temperature 
distribution on the surface of the star, with small, hot regions at the magnetic poles. Highly
modulated thermal emission with high temperatures will then be produced. These
results have been applied to model the observed emission from RXJ 0720.4--3125, 
a highly magnetized neutron star, with apparent success \citep{ppmm06}. Pulsed fractions as 
high as $\sim$30\% in the case of isotropic blackbody emission have been reported 
\citep{gkp05}. Therefore, it remains to 
be shown whether the same models can be applied to reproduce the observed emission 
characteristics in \psr. In this case, \psr\ would be the first radio pulsar to show the
effects of a high magnetic field through its X-ray emission.

We also point 
out the thermal emission with high temperature and high pulsed fraction that was 
found for PSR J1852+0040 \citep{ghs05}. Although a detailed timing solution has not been 
reported, initial estimates suggest a characteristic age of $\tau_c$ $>$ 24 kyr and low 
magnetic field of $B$ $<$ 3$\times$10$^{12}$~G. If these estimates are correct, existing theories 
for thermal emission from neutron stars cannot readily account for the observed characteristics, 
including those involving high-magnetic-field effects as mentioned above.

\subsection{Other High-Magnetic Field Pulsars} \label{ssec:other}
Many radio pulsars having inferred magnetic fields in the range 10$^{13-14}$~G have now 
been discovered. A sample of these pulsars with associated X-ray observations is shown in
Table \ref{tabPsrs}. Most of these sources have proved to be very faint in X-rays. Only
two pulsars, PSRs J1846--0258 and B1509--58, are bright non-thermal sources and 
power bright PWNe. As expected, and in agreement with normal radio pulsars, they are 
young and very energetic ($\tau_c$ $<$ 2,000 yrs and $\dot{E}$ $>$ 10$^{36}$ ergs s$^{-1}$).

On the other hand, the older and less energetic pulsars in Table \ref{tabPsrs} have not been
detected in X-rays ($\tau_c$ $>$ 10,000 yrs and $\dot{E}$ $<$ 10$^{33}$ ergs s$^{-1}$). This 
includes PSR J1847--0130, the radio pulsar with highest inferred magnetic field discovered 
to date (0.9$\times$10$^{14}$~G). These pulsars then show no enhancement of high-energy 
emission despite having inferred magnetic fields in the magnetar range. One intermediate case 
is that of PSR J1718--3718, which does have 
a faint X-ray counterpart seen with \chan. However, the detailed characteristics of this emission
(e.g., thermal vs. non-thermal) could not be constrained with the data.

\psr\ is then an interesting
and puzzling source. Despite being young and energetic, it does not power a bright PWN 
and we have found its emission to be dominated by a thermal component. This is in direct 
contrast to the sources mentioned above, particularly PSR J1846--0258 with which \psr\ 
shares almost identical spin characteristics and even similar surroundings in their respective
SNRs \citep{hcg03,gs05}. It is therefore unclear what the
physical reasons are behind their vastly different X-ray emission. In addition, while
it is possible that the characteristics observed in \psr\ may be due to heat conductivity
effects on a highly magnetized atmosphere, the emission is not magnetar-like.

\section{Conclusion}\label{sec:con}
The X-ray emission from the young, high magnetic field radio pulsar \psr\ shows a thermal 
spectrum with high temperature and small emitting radius, making it the radio pulsar with 
smallest characteristic age from which thermal X-ray emission has been detected. The pulse 
profile of this emission 
is consistent with a single, narrow pulse with a high pulsed fraction. Hot spots heated by 
back-flowing particles from the magnetosphere are not expected in such a young source,
while the X-ray characteristics are not consistent with cooling emission from the whole surface.
However, a highly anisotropic temperature distribution on the surface due to a high 
magnetic field may be able to account for the observed characteristics. This would make \psr\
the first radio pulsar to exhibit high-magnetic-field effects on its X-ray emission. Additional
X-ray observations, particularly with improved temporal resolution, will help to confirm and 
constrain the observed characteristics.

Many high magnetic field radio pulsars have now been observed 
in X-rays, some with very similar spin characteristics to \psr\ and others with higher inferred 
fields, but it remains unclear why \psr\ is to date the only one to show such effects.
Recently, the discovery 
of radio emission from a magnetar \citep{crh+06} has shown that such emission is possible
in these sources, contributing to our understanding of the mechanisms at work. However, we 
still lack an understanding for the absence of magnetar-like emission from radio 
pulsars with high magnetic fields and additional observations of these objects 
are then needed.



\bibliography{journals1,psrrefs,modrefs}   

\end{document}